\documentstyle[prd,aps,preprint,tighten,epsfig]{revtex}

\begin{document}

\draft

\title{Democratic Neutrino Mixing Reexamined}
\author{\bf Harald Fritzsch}
\address{Sektion Physik, Universit$\ddot{a}$t M$\ddot{u}$nchen,
Theresienstrasse 37A, 80333 Munich, Germany}
\author{\bf Zhi-zhong Xing}
\address{Institute of High Energy Physics, Chinese Academy of Sciences,
Beijing 100039, China \\
({Electronic address: xingzz@mail.ihep.ac.cn}) } \maketitle

\begin{abstract}
We reexamine the democratic neutrino mixing ansatz, in which the
mass matrices of charged leptons and Majorana neutrinos arise
respectively from the explicit breaking of $\rm S(3)_L \times
S(3)_R$ and $\rm S(3)$ flavor symmetries. It is shown that a
democracy term in the neutrino sector can naturally
allow the ansatz to fit the solar neutrino mixing angle
$\theta_{\rm sun} \approx 33^\circ$. We predict $\sin^2
2\theta_{\rm atm} \approx 0.95$ for atmospheric neutrino
mixing and ${\cal J} \approx 1.2\%$ for leptonic CP
violation in neutrino oscillations without any fine-tuning.
Direct relations between the model parameters and experimental
observables are also discussed.
\end{abstract}


The recent solar \cite{SNO}, atmospheric \cite{SK}, KamLAND
\cite{KM} and K2K \cite{K2K} neutrino oscillation experiments
provide us with very compelling evidence that neutrinos are
massive and lepton flavors are mixed. To account for the observed
neutrino mass-squred differences ($\Delta m^2_{\rm sun} \sim 6.9
\times 10^{-5} ~ {\rm eV}^2$ and $\Delta m^2_{\rm atm} \sim 2.3
\times 10^{-3} ~ {\rm eV}^2$) and mixing factors ($\sin^2
2\theta_{\rm sun} \sim 0.84$ and $\sin^2 2\theta_{\rm atm} \sim
1.0$ \cite{Valle}), many phenomenological models of
lepton mass matrices have been proposed in the literature
\cite{Review}. Some of them take advantage of the idea of flavor
democracy, from which the largeness of two lepton mixing angles,
the smallness of three quark mixing angles, and the wide mass gaps
between $(m_\tau, m_t, m_b)$ and their lighter counterparts can
simultaneously be understood.

The original ansatz of democratic neutrino mixing \cite{FX96} is based
on the phenomenological conjecture that charged lepton and Majorana
neutrino mass matrices may arise from the breaking of
$\rm S(3)_{\rm L}\times S(3)_{\rm R}$ and $\rm S(3)$ flavor symmetries,
respectively:
\begin{eqnarray}
M_l & = & \frac{c^{~}_l}{3} \left (\matrix{
1       & 1     & 1 \cr
1       & 1     & 1 \cr
1       & 1     & 1 \cr} \right ) + \Delta M_l \; ,
\nonumber \\
M_\nu & = & c_\nu \left (\matrix{
1       & 0     & 0 \cr
0       & 1     & 0 \cr
0       & 0     & 1 \cr} \right ) + \Delta M_\nu \; ,
%       (1)
\end{eqnarray}
where $c^{~}_l$ and $c_\nu$ measure the
corresponding mass scales of charged leptons and neutrinos.
The explicit symmetry-breaking term $\Delta M_l$
is responsible for the generation of muon and electron masses,
and $\Delta M_\nu$ is responsible for the breaking of neutrino
mass degeneracy. A very simple form of $\Delta M_l$ and
$\Delta M_\nu$ reads \cite{FX96}
\begin{eqnarray}
\Delta M_l & = & \frac{c^{~}_l}{3} \left ( \matrix{
-i\delta_l       & 0     & 0 \cr
0       & i\delta_l     & 0 \cr
0       & 0     & \varepsilon^{~}_l \cr } \right ) \; ,
\nonumber \\
\Delta M_\nu & = & c_\nu \left ( \matrix{
-\delta_\nu       & ~ 0     & 0 \cr
0       & ~ \delta_\nu     & 0 \cr
0       & ~ 0     & \varepsilon_\nu \cr } \right ) \; ,
%       (2)
\end{eqnarray}
where $(\delta_l, \varepsilon^{~}_l)$ and
$(\delta_\nu, \varepsilon_\nu)$ are real dimensionless perturbation
parameters of small magnitude, and the imaginary phase of
$\Delta M_l$ is a natural source of leptonic CP violation in
neutrino oscillations. Because $M_\nu$ is already diagonal, we only
need to diagonalize $M_l$ by means of the orthogonal transformation
$V M_l V^T = {\rm Diag}\{m_e, m_\mu, m_\tau\}$,
in order to express the leptonic charged-current interactions in
terms of the mass eigenstates of charged leptons and neutrinos.
The lepton flavor mixing matrix is just given by the unitary
matrix $V$; i.e.,
\begin{equation}
V \approx \left ( \matrix{
\frac{1}{\sqrt{2}} & \frac{-1}{\sqrt{2}} & 0 \cr\cr
\frac{1}{\sqrt{6}} & \frac{1}{\sqrt{6}} & \frac{-2}{\sqrt{6}} \cr\cr
\frac{1}{\sqrt{3}} & \frac{1}{\sqrt{3}} & \frac{1}{\sqrt{3}} \cr}
\right ) +
i \sqrt{\frac{m_e}{m_\mu}}
\left ( \matrix{
\frac{1}{\sqrt{6}}      & ~ \frac{1}{\sqrt{6}} ~        &
\frac{-2}{\sqrt{6}} \cr\cr
\frac{1}{\sqrt{2}}      & ~ \frac{-1}{\sqrt{2}} ~       & 0 \cr\cr
0       & ~ 0 ~ & 0 \cr} \right )
+ \frac{m_\mu}{m_\tau} \left ( \matrix{
0       & 0     & 0 \cr\cr
\frac{1}{\sqrt{6}}      & \frac{1}{\sqrt{6}}    & \frac{1}{\sqrt{6}} \cr\cr
\frac{-1}{\sqrt{12}}    & \frac{-1}{\sqrt{12}}  & \frac{1}{\sqrt{3}}
\cr} \right ) \; .
%       (3)
\end{equation}
Given $m_e/m_\mu \approx 0.00484$ and $m_\mu/m_\tau \approx
0.0594$ \cite{PDG}, the mixing factors of solar and atmospheric
neutrino oscillations turn out to be
\begin{eqnarray}
\sin^2 2\theta_{\rm sun} & \approx &
1 - \frac{4}{3} \frac{m_e}{m_\mu} \; \approx \; 0.99 \; ,
\nonumber \\
\sin^2 2\theta_{\rm atm} & \approx &
\frac{8}{9} \left ( 1+ \frac{m_\mu}{m_\tau} \right )
\; \approx \; 0.94 \; .
%       (4)
\end{eqnarray}
The result of $\sin^2 2\theta_{\rm sun}$ is obviously disfavored by
current solar neutrino data, and that of $\sin^2 2\theta_{\rm atm}$
apparently deviates from the maximal atmospheric neutrino mixing.

A simple way to suppress the afore-obtained value of
$\sin^2 2\theta_{\rm sun}$ and enhance that of
$\sin^2 2\theta_{\rm atm}$ is to add another S(3)-symmetry term,
which was not included in Eq. (1), into the neutrino mass
matrix $M_\nu$ \cite{Tanimoto}. In this case, we have
\begin{equation}
M_\nu \; = \; c_\nu \left [ \left (\matrix{
1       & 0     & 0 \cr
0       & 1     & 0 \cr
0       & 0     & 1 \cr} \right ) +
r_\nu \left (\matrix{
1       & 1     & 1 \cr
1       & 1     & 1 \cr
1       & 1     & 1 \cr} \right ) \right ] + \Delta M_\nu \; ,
%       (5)
\end{equation}
where $r_\nu$ is in principle an arbitrary parameter. To get large
lepton mixing angles, however, $|r_\nu| \ll 1$ must be satisfied.
It is shown in Ref. \cite{Tanimoto} that the $r_\nu$-induced
corrections to $\sin^2 2\theta_{\rm sun}$ and
$\sin^2 2\theta_{\rm atm}$ can both be constructive, and
$\Delta m^2_{\rm sun} \sim (1 - 2) \times 10^{-4} ~ {\rm eV}^2$
is predicted by taking the appropriate parameter space
of $(c_\nu, r_\nu, \delta_\nu, \varepsilon_\nu)$
%%%%%%%%%%%%%%%%%%%%%%%%%%%%%
\footnote{Note that the diagonal perturbation term of $M_\nu$ in
Ref. \cite{Tanimoto} is not exactly the same as our $\Delta M_\nu$
given in Eq. (2).}.
%%%%%%%%%%%%%%%%%%%%%%%%%%%%%
Note that
$\Delta m^2_{\rm sun} \sim {\cal O}(10^{-4}) ~ {\rm eV}^2$
is no more favored by today's experimental data on solar neutrino
oscillations. It is therefore necessary to reexamine whether a
favorable bi-large neutrino mixing pattern can naturally be derived
from the explicit breaking of $\rm S(3)_{\rm L}\times S(3)_{\rm R}$
symmetry of charged leptons and $\rm S(3)$ symmetry of Majorana
neutrinos. If the answer remains affirmative, then direct and testable
relations between the model parameters and experimental observables
should be established.

The main purpose of this short paper is to demonstrate that the
$r_\nu$-modified version of our phenomenological ansatz is
actually compatible with current neutrino oscillation data. We
find that the experimentally-favored value of
$\sin^2 2\theta_{\rm sun}$ can
naturally be achieved. We derive a simple relation between $\sin^2
2\theta_{\rm atm}$ and $\cos 2\theta_{\rm sun}$, and then arrive
at the prediction $\sin^2 2\theta_{\rm atm} \approx 0.95$ without
any fine-tuning. We also show how to relate the model
parameters to the relevant observables. Our analytical results
will be very useful to test the democratic neutrino mixing
scenario, when more accurate experimental data are available in
the near future.

For simplicity, we take $c_\nu$, $r_\nu$, $\delta_\nu$ and
$\varepsilon_\nu$ in Eq. (5) to be real and positive. Then
$M_\nu$ can be diagonalized by means of a real orthogonal
transformation $U^T M_\nu U = {\rm Diag}\{m_1, m_2, m_3\}$.
It is obvious that three neutrino masses must be nearly
degenerate. Taking the convention $m_1 < m_2 < m_3$, we obtain
\begin{eqnarray}
m_1 & \approx & c_\nu \left ( 1 + r_\nu - \sqrt{r^2_\nu +
\delta^2_\nu} \right ) \; ,
\nonumber \\
m_2 & \approx & c_\nu \left ( 1 + r_\nu + \sqrt{r^2_\nu +
\delta^2_\nu} \right ) \; ,
\nonumber \\
m_3 & \approx & c_\nu \left ( 1 + r_\nu + \varepsilon_\nu \right ) \; .
%       (6)
\end{eqnarray}
The near degeneracy of three neutrino masses implies that the
effective mass-squared term of the tritium beta decay, defined as
$\langle m\rangle^2_e \equiv \sum (m^2_i |V_{ei}|^2)$ for $i=1,2$
and $3$, approximately amounts to $c^2_\nu$. In other words, $c_\nu
\approx \langle m\rangle_e$ holds. Then we obtain $c_\nu < 2.2$ eV
from the direct-mass-search experiments \cite{PDG} and $c_\nu <
0.23$ eV from the recent WMAP observational data \cite{WMAP}.
In view of $\Delta m^2_{\rm atm} \gg \Delta m^2_{\rm sun}$, we require
$\varepsilon_\nu \gg r_\nu$ and $\varepsilon_\nu \gg \delta_\nu$.
Therefore,
\begin{eqnarray}
\Delta m^2_{\rm sun} & = & \Delta m^2_{21} \approx
4 c^2_\nu \sqrt{r^2_\nu + \delta^2_\nu} \; ,
\nonumber \\
\Delta m^2_{\rm atm} & = & \Delta m^2_{32} \approx
2 c^2_\nu \varepsilon_\nu \; .
%       (7)
\end{eqnarray}
As for the orthogonal matrix $U$, its nine elements $U_{1i}$, $U_{2i}$
and $U_{3i}$ (for $i=1,2,3$) have the following relations:
\begin{eqnarray}
U_{2i} & = & \frac{c_\nu \left ( 1 - \delta_\nu \right ) - m_i}
{c_\nu \left ( 1 + \delta_\nu \right ) - m_i} U_{1i} \; ,
\nonumber \\
U_{3i} & = & \frac{ -r_\nu \left ( U_{1i} + U_{2i} \right )}
{c_\nu \left ( 1 + r_\nu + \varepsilon_\nu \right ) - m_i} \; .
%       (8)
\end{eqnarray}
To the accuracy of ${\cal O}(r_\nu/\varepsilon_\nu)$, the
expression of $U$ is found to be
\begin{equation}
U \; \approx \; \left ( \matrix{
\cos\theta & \sin\theta & \displaystyle\frac{r_\nu}{\varepsilon_\nu} \cr
-\sin\theta & \cos\theta & \displaystyle\frac{r_\nu}{\varepsilon_\nu} \cr
\displaystyle\frac{r_\nu}{\varepsilon_\nu} \left ( \sin\theta -
\cos\theta \right )
& \displaystyle -\frac{r_\nu}{\varepsilon_\nu} \left ( \sin\theta +
\cos\theta \right ) & 1 \cr} \right ) \; ,
%       (9)
\end{equation}
where $\tan 2\theta \equiv r_\nu/\delta_\nu$. Without loss of
generality, $\theta$ is required to lie in the first quadrant. It
is clear that $U$ becomes the unity matrix in the limit $r_\nu
=0$. In the case of $r_\nu \neq 0$, the lepton flavor mixing
matrix takes the form $\widehat{V} = V U$, where $V$ has been
given in Eq. (3). We explicitly obtain $\widehat{V} \approx
\widehat{V}_0 + \widehat{V}_1$ as a good approximation, in which
\begin{equation}
~ \widehat{V}_0 \; = \; \left ( \matrix{ \frac{1}{\sqrt{2}} \left
( \cos\theta + \sin\theta \right ) & ~ \frac{-1}{\sqrt{2}} \left (
\cos\theta - \sin\theta \right ) ~ & 0 \cr\cr \frac{1}{\sqrt{6}}
\left ( \cos\theta - \sin\theta \right ) & ~ \frac{1}{\sqrt{6}}
\left ( \cos\theta + \sin\theta \right ) ~ & \frac{-2}{\sqrt{6}}
\cr\cr \frac{1}{\sqrt{3}} \left ( \cos\theta - \sin\theta \right )
& ~ \frac{1}{\sqrt{3}} \left ( \cos\theta + \sin\theta \right ) ~
& \frac{1}{\sqrt{3}} \cr} \right ) \; ,
%       (10)
\end{equation}
and
\begin{eqnarray}
\widehat{V}_1 & = & i \sqrt{\frac{m_e}{m_\mu}} \left ( \matrix{
\frac{1}{\sqrt{6}} \left ( \cos\theta - \sin\theta \right ) & ~
\frac{1}{\sqrt{6}} \left ( \cos\theta + \sin\theta \right ) &
\frac{-2}{\sqrt{6}} \cr\cr \frac{1}{\sqrt{2}} \left ( \cos\theta +
\sin\theta \right ) & ~ \frac{-1}{\sqrt{2}} \left ( \cos\theta -
\sin\theta \right )  & 0 \cr\cr 0       & 0     & 0 \cr} \right )
~~
\nonumber \\
&  & + ~ \frac{m_\mu}{m_\tau} \left ( \matrix{ 0       & 0     & 0
\cr\cr \frac{1}{\sqrt{6}} \left ( \cos\theta - \sin\theta \right )
& \frac{1}{\sqrt{6}} \left ( \cos\theta + \sin\theta \right ) &
\frac{1}{\sqrt{6}} \cr\cr \frac{-1}{\sqrt{12}} \left ( \cos\theta
- \sin\theta \right ) & \frac{-1}{\sqrt{12}} \left ( \cos\theta +
\sin\theta \right ) & \frac{1}{\sqrt{3}} \cr} \right )
\nonumber \\
&  & + ~ \frac{r_\nu}{\varepsilon_\nu} ~ \left ( \matrix{
0 & 0 & 0 \cr\cr
\frac{2}{\sqrt{6}} \left ( \cos\theta - \sin\theta \right )
& ~ \frac{2}{\sqrt{6}} \left ( \cos\theta + \sin\theta \right ) ~
& \frac{2}{\sqrt{6}} \cr\cr
\frac{-1}{\sqrt{3}} \left ( \cos\theta - \sin\theta \right )
& ~ \frac{-1}{\sqrt{3}} \left ( \cos\theta + \sin\theta \right ) ~
& \frac{2}{\sqrt{3}} \cr}
\right ) \; .
%       (11)
\end{eqnarray}
Comparing $\widehat{V}$ with $V$, we see that
$\widehat{V}_{e3} \approx V_{e3}$ holds. This result implies that
the mixing angle $\theta_{13}$ in the standard parametrization of
$\widehat{V}$ \cite{PDG} is rather small:
\begin{equation}
\sin\theta_{13} \; = \; |\widehat{V}_{e3}| \approx
\frac{2}{\sqrt{6}} \sqrt{\frac{m_e}{m_\mu}} \approx 0.057 \; ,
%       (12)
\end{equation}
or $\theta_{13} \approx 3.2^\circ$. On the other hand, eight other
elements of $\widehat{V}$ may get appreciable $r_\nu$-induced
corrections.

With the help of Eqs. (10) and (11), the solar neutrino mixing
factor is obtained as
\begin{equation}
\sin^2 2\theta_{\rm sun} \; = \; 4 |\widehat{V}_{e1}|^2
|\widehat{V}_{e2}|^2 \; \approx \; \cos^2 2\theta \; .
%       (13)
\end{equation}
It follows that $\theta \approx (45^\circ - \theta_{\rm sun})$
holds. In other words, $\theta$ measures the deviation of
$\theta_{\rm sun}$ from $45^\circ$. As observed in Ref.
\cite{Raidal}, the sum $\theta_{\rm sun} + \theta_{\rm C} \approx
45^\circ$ with $\theta_{\rm C}$ being the Cabibbo angle of quark
mixing is favored by current experimental data. In this case, we
are then left with $\theta \approx \theta_{\rm C} \approx
12^\circ$. The ratio $r_\nu/\delta_\nu$ can in turn be determined
in terms of the mixing angle $\theta_{\rm sun}$: $r_\nu/\delta_\nu
\approx \cot 2\theta_{\rm sun}$. Typically taking the best-fit
value $\theta_{\rm sun} \approx 33^\circ$, we arrive at
$r_\nu/\delta_\nu \approx 0.44$. One may also estimate the
magnitude of $r_\nu/\varepsilon_\nu$ with the help of Eq. (7). The
result is
\begin{equation}
\frac{r_\nu}{\varepsilon_\nu} \approx \frac{\Delta m^2_{\rm
sun}}{\Delta m^2_{\rm atm}} \cdot \frac{\cos 2\theta_{\rm sun}}{2}
\approx 6.1 \times 10^{-3} \; ,
%       (14)
\end{equation}
where $\Delta m^2_{\rm sun}/\Delta m^2_{\rm atm} \approx 3 \times
10^{-2}$ and $\theta_{\rm sun} \approx 33^\circ$ have
been used. It is then clear that $\varepsilon_\nu \gg \delta_\nu
\sim r_\nu$ holds.

Now let us calculate the atmospheric neutrino mixing factor
$\sin^2 2\theta_{\rm atm}$ by using Eqs. (10) and (11). We obtain
\begin{equation}
\sin^2 2\theta_{\rm atm} = 4 |\widehat{V}_{\mu 3}|^2 \left ( 1 -
|\widehat{V}_{\mu 3}|^2 \right ) \approx \frac{8}{9} \left ( 1+
\frac{m_\mu}{m_\tau} + \frac{\Delta m^2_{\rm sun}}{\Delta m^2_{\rm
atm}} \cos 2\theta_{\rm sun} \right ) \approx 0.95 \; .
%       (15)
\end{equation}
Comparing between Eqs. (4) and (15), we find that the
$r_\nu$-induced correction to $\sin^2 2\theta_{\rm atm}$ is
constructive but suppressed by $\Delta m^2_{\rm sun}/\Delta
m^2_{\rm atm} \sim {\cal O}(10^{-2})$. We conclude that the
maximal atmospheric neutrino mixing cannot be achieved in a simple
and natural way, unless the ratio $\Delta m^2_{\rm sun}/\Delta
m^2_{\rm atm}$ is as large as of ${\cal O}(10^{-1})$. A more
precise determination of $\Delta m^2_{\rm sun}$, $\Delta m^2_{\rm
atm}$, $\theta_{\rm sun}$ and $\theta_{\rm atm}$ will test the
validity of Eq. (15).

The consequences of this phenomenological ansatz on the
neutrinoless double beta decay and CP violation in neutrino
oscillations are interesting. A straightforward calculation yields
$\langle m\rangle_{ee} \equiv | \sum (m_i \widehat{V}^2_{ei})|
\approx c_\nu$ for the effective mass of the neutrinoless double
beta decay. It becomes obvious that $\langle m\rangle_{ee} \approx
\langle m\rangle_e \approx c_\nu$ holds. The absolute
neutrino mass scale in our ansatz can be fixed either from a
measurement of the tritium beta decay or from a positive signal of
the neutrinoless double beta decay. The Jarlskog invariant of
CP violation \cite{J} is found to be
\begin{equation}
{\cal J} \; \approx \; \frac{1}{3\sqrt{3}}
\sqrt{\frac{m_e}{m_\mu}} \sin 2 \theta_{\rm sun} \; \approx \;
0.012 \; .
%       (16)
\end{equation}
Such a strength of leptonic CP violation is likely to be observed
in a long-baseline neutrino oscillation experiment.

One can see that the democratic neutrino mixing ansatz under
discussion is compatible with all of current neutrino data. Its
prediction for $\sin^2 2\theta_{\rm atm}$, which apparently
deviates from the maximal atmospheric neutrino mixing, can easily
be tested in the near future. Of course, part of our results
depend on the explicit symmetry breaking patterns (i.e., $\Delta
M_l$ and $\Delta M_\nu$). Let us comment on the effects of $\rm
S(3)$ flavor symmetry breaking terms in some detail:

(a) The lepton flavor mixing matrix $\widehat{V}$ is insensitive
to the form of $\Delta M_l$, as already observed in Ref. \cite{FX96}.
The point is simply that the strong mass
hierarchy of three charged leptons makes the contribution of
$\Delta M_l$ to $\widehat{V}$ insignificant, no matter whether
$\Delta M_l$ is diagonal or off-diagonal.

(b) If a contrived and fine-tuned pattern of $\Delta M_\nu$ is
taken, it should be possible to obtain a ``proper'' (2,3)-rotation
angle from $M_\nu$ in order to arrive at
$\theta_{\rm atm} \sim 45^\circ$. However, it is more natural
to consider the simple forms of $\Delta M_\nu$ such as the
diagonal perturbation given in Eq. (2), at least from the point of
view of model building \cite{Koide}.

(c) A remarkable advantage of the diagonal perturbation
$\Delta M_\nu$ is that it guarantees $M_\nu$ to be stable against
radiative corrections \cite{Tanimoto,Xing01}, although three mass
eigenvalues of $M_\nu$ are almost degenerate. This feature makes
sense for model building too, because the
$\rm S(3)_L \times S(3)_R$ symmetry of $M_l$ and the
S(3) symmetry of $M_\nu$ are most likely to manifest themselves
at a high energy scale (e.g., the seesaw scale \cite{SS}, where
three heavy right-handed neutrinos might also have an approximate
flavor democracy \cite{Branco}).

Finally, it is worth emphasizing that four free parameters of
$M_\nu$ may all be determined in terms of the relevant observable
quantities. We obtain $c_\nu \approx \langle m\rangle_e \approx
\langle m\rangle_{ee}$. Then $\varepsilon_\nu \approx \Delta
m^2_{\rm atm}/(2\langle m\rangle^2_e)$ can straightforwardly be
derived from Eq. (7). With the help of Eq. (14), we further arrive
at
\begin{eqnarray}
r_\nu & ~ \approx ~ & \frac{\Delta m^2_{\rm sun}}{\langle m\rangle^2_e}
\cdot \frac{\cos 2\theta_{\rm sun}}{4} \; ,
\nonumber \\
\delta_\nu & ~ \approx ~ & \frac{\Delta m^2_{\rm sun}}
{\langle m\rangle^2_e} \cdot \frac{\sin 2\theta_{\rm sun}}{4} \; .
%       (17)
\end{eqnarray}
Note that the magnitudes of $\varepsilon_\nu$ and $\delta_\nu$
should be of or below ${\cal O}(0.1)$, because they are perturbative
parameters of $\Delta M_\nu$. Taking $\varepsilon_\nu \sim 0.1$,
for instance, we may get $\langle m\rangle_e \approx
\langle m\rangle_{ee} \sim 0.1$ eV. To measure such a small
$\langle m\rangle_e$ in the tritium beta decay is practically difficult
(but not impossible) in the near future \cite{T}.
In comparison, $\langle m\rangle_{ee} \sim 0.1$ eV is definitely
accessible in a number of planned experiments for the neutrinoless
double beta decay \cite{2B}. This numerical example indicates that
$\varepsilon_\nu \sim {\cal O}(0.1)$ is most plausible. A much smaller
$\varepsilon_\nu$ would make $m_i \approx c_\nu$ (for $i=1,2,3$)
too large to be compatible with the WMAP upper limit on
$m_i$, while a much bigger $\varepsilon_\nu$ would loss its physical
meaning as a perturbative parameter.

In summary, we have reexamined the democratic neutrino mixing
ansatz by taking into account an extra S(3)-symmetry term in the
Majorana neutrino mass matrix. After explicit symmetry breaking
induced by the diagonal perturbations, we obtain the mass spectrum
of charged leptons with a strong hierarchy and that of neutrinos
with a near degeneracy. The suppressed democracy term in the
neutrino sector can naturally permit the model to fit current
solar neutrino oscillation data with $\theta_{\rm sun} \approx
33^\circ$. We have derived a simple relation between $\sin^2
2\theta_{\rm atm}$ and $\cos 2\theta_{\rm sun}$, and achieved the
prediction $\sin^2 2\theta_{\rm atm} \approx 0.95$ without any
fine-tuning. Whether this atmospheric neutrino mixing factor is
really maximal or not will provide a sensitive test of our
phenomenological ansatz. We have also established the direct
relations between the model parameters and relevant experimental
observables. We remark that the democratic neutrino mixing
scenario is simple, viable and suggestive. It could be useful for
model building, in particular at a high energy scale at which the
$\rm S(3)_L \times S(3)_R$ symmetry of charged leptons and the
S(3) symmetry of Majorana neutrinos are expected to become
relevant.

\vspace{0.5cm}

One of us (Z.Z.X.) is grateful to S. Zhou for useful discussions
and partial involvement at the early stage of this work, which was
supported in part by the National Natural Science Foundation of
China.

\end{document}